# Resonant Low Frequency Interlayer Shear Modes in Folded Graphene Layers

Chunxiao Cong[1] and Ting Yu[1,2,3*]

[1]Division of Physics and Applied Physics, School of Physical and Mathematical Sciences, Nanyang Technological University, 637371, Singapore;

[2]Department of Physics, Faculty of Science, National University of Singapore, 117542, Singapore;

[3]Graphene Research Center, National University of Singapore, 117546, Singapore

*Address correspondence to yuting@ntu.edu.sg

**Naturally or artificially stacking extra layers on single layer graphene (SLG) forms few-layer graphene (FLG), which has attracted tremendous attention owing to its exceptional properties inherited from SLG and new features generated by introducing extra freedom. In FLG, shear modes play a critical role in understanding its distinctive properties. Unfortunately, energies of shear modes are so close to excitation lasers to be fully blocked by a Rayleigh rejecter. This greatly hinders investigations of shear modes in FLG. Here, we demonstrate dramatically enhanced shear modes in properly folded FLG. Benefiting from the extremely strong signals, for the first time, enhancement mechanism, vibrational symmetry, anharmonicity and electron-phonon coupling (EPC) of shear modes are uncovered through studies of two-dimensional (2D) Raman mapping, polarization- and temperature-dependent Raman spectroscopy. This work complements Raman studies of graphene layers, and paves an efficient way to exploit low frequency shear modes of FLG and other 2D layered materials.**

Few-layer graphene (FLG) possesses unique properties of crystal structure, lattice dynamics and electronics, for example, opening an energy band gap in Bernal-stacked bilayer graphene (BLG)[1]; different responses in the integer quantum Hall effect (IQHE) measurements for ABA- and ABC-stacked trilayer graphene (TLG)[2]; and the formation of Van Hove Singularity (VHS) in folded or twisted double layer graphene (f/tDLG)[3-4]. Raman spectroscopy is one of the most useful and versatile techniques to probe graphene layers as has been demonstrated in the studies of number of layers, strain, doping, edges, stacking orders, and even magneto-phonon coupling in graphene layers[5-20]. The beauty of Raman spectroscopy for exploiting graphene is the truth that the fundamental vibrational modes such as G, G' and D modes are resonant with electrons, which leads to very strong signal, facilitates many unfulfillable measurements with weak Raman signal and provides an effective way to probe phonons and electronic band structures through strong electron-phonon coupling (EPC). In addition to these well-known fundamental modes, some other weak modes like higher order, combinational and superlattice wave vector mediated phonon modes have been observed in either Bernal- or nonBernal-stacked graphene layers[21-24]. They all carry interesting and important information about lattice vibration and electronic band structures.

Another very fundamental and intrinsic vibrational mode in FLG and bulk graphite is rigid interlayer shear mode, involving the relative motion of atoms in adjacent layers. Vibrational energies of shear modes vary when the thickness and consequently the restoring force strength of Bernal-stacked graphene layers changes as being demonstrated by the experimental observation and perfectly modeled by a simple linear chin system[25]. Therefore, this shear mode, named C peak can be used as another Raman spectroscopic feature for identifying the thickness of Bernal-stacked graphene layers. Considering its low energy, ~ 5 meV, researchers believe the C peak could be a probe for the quasiparticles near the Dirac point through quantum interference[25]. However, the low energy also causes direct observation of shear modes being extremely challenging because the shear modes are so close to the excitation photons and fully suppressed by a notch or edge filter of most Raman instruments. To directly detect this C peak, low-doped Si substrate with pre-etched holes was used in the previous study[25]. Though the C peak of the suspended graphene layers was observed on such specially prepared substrate, it is still very weak, especially for BLG and TLG, which happen to be the most interesting and promising candidates of graphene family together with SLG. Therefore, the extremely weak signal and the sophisticated sample preparation severely limit study of shear mode and its coupling with other particles. After the pioneering work[25], very few experimental observations of the first order fundamental shear modes of FLG were reported[26].

Here, we report remarkably enhanced shear modes in folded 2+2 and 3+3 graphene layers with some certain rotational angles, where VHS is induced and results in the enhancement and doublet splitting of G mode as discussed in our previous study[27]. These folded 2+2 and 3+3 graphene layers with strongly enhanced G mode are named as $r$-f4LG and $r$-f6LG, respectively. Here, "$r$" refers to resonance of G mode. Instead of specially prepared substrate of low-doped Si with array of micro-holes[25], we used the typical high-doped Si with $SiO_2$ (285 nm) substrate. The extremely strong signal, comparable or even stronger than the resonant G mode, enables measurements of two-dimensional Raman mapping, polarization- and temperature-dependent Raman spectroscopy of this low frequency shear mode for the first time and thus unravels its vibrational symmetry, anharmonicity and EPC.

**Results**

In our previous study, we classified folded graphene layers into three types by the folding or rotational angles θ of $\theta_{small}$, $\theta_{medium}$ and $\theta_{large,}$ for a given excitation laser. These three types of folded layers exhibit very different Raman spectral features[27]. Figure 1 shows BLG with two self-folded regions of rotational angles of 11º ($\theta_{medium}$) and 21.4º ($\theta_{large,}$). Excitation laser of 532 nm was used for all the Raman measurements in this work. The significant enhancement of G mode in the $\theta_{medium}$ *r*-f4LG can be clearly seen in the Raman image (Fig. 1b) and spectra (Fig. 1g). Surprisingly, along with the resonant G mode, a low frequency (~30 cm$^{-1}$) peak also presents in this *r*-f4LG and it is so strong that the *first* Raman images of such low frequency mode in graphene layers are clearly resolved by extracting its intensity, position and width (Fig. 1d-f). A single Lorentzian lineshape peak is used for fitting the Raman peaks in this study (Fig. S1). Reading the position and width of this low frequency mode and referring to the previous study of the suspended BLG[25], we tentatively assign it to the interlayer shear mode (C peak or our label, $C_2$) of the "mother" flake, BLG, but with an extremely large enhancement of the intensity. The good correlation between the enhanced $C_2$ and G modes is clearly revealed by the Raman images (Fig. 1b and d) and spectrum (Fig. 1g). Therefore, we believe that they share the same enhancement mechanism, a folding induced VHS. Though the formation and influence of VHS in 1+1 f/tDLG has been intensively studied recently[3-4], there is no evidence of the existence of such VHS in 2+2 $\theta_{medium}$ *r*-f4LG. In this work, by adapting the previous methodology[28-29], we exploit the electronic band structure and density of states (DOS) of such 2D system. The VHS with good corresponding to our excitation photon energy is seen (Fig. S2). We remark that the nonuniformity appearing in the Raman images of intensity of $C_2$ mode (Fig. 1d) might be due to the varying of the interlayer spacing. A relatively weak peak locating at around 115 cm$^{-1}$ is noticed and attributed to a combinational mode of interlayer breathing mode (~86 cm$^{-1}$)[30] and shear mode, labelled as B+C$_2$ peak herein.

Mediated by either the short-range twisted bilayer lattice or the supperlattice, ZO' and *R* peaks have also been observed in 1+1 *r*-fDLG and exhibit dependence of peak positions on twisting angles[23,31]. To further prove this ultra-strong low frequency peak is the shear mode of BLG and unravel its nature, we measured other two pieces of *r*-f4LGs, which present *R* peaks of various positions (Fig. 1h). The rotational angles are determined by carefully fitting and reading the positions of *R* peaks[23-24]. Clearly enough, all C$_2$ peaks are enhanced and their positions show no dependence on the rotational angles, which is very much different to the twisting angle dependent peaks discussed previously. The detailed curve fitting and calculation (Table S1) show the peak positions and linewidths of these C$_2$ peaks as well as the derived interlayer coupling strength by the frequencies of the C$_2$ peaks. The derived interlayer coupling strength keeps the same as that of normal Bernal-stacked BLG[25]. This indicates this fundamental shear phonon mode of the "mother" flake, though could be enhanced by the folding, is very robust even after being folded on top of itself, which must be very interesting and important for exploiting mechanical and electrical properties of such folded atomically thin layers including graphene and other 2D systems. It is also noticed that the position of the weak combinational (B+C$_2$) peak does not change neither when the rotational angles vary. This further supports our assignment since both breathing (B) and shear (C$_2$) modes are the fundamental modes of the "mother" flake and are not affected by the folding.

Not only the 2+2 f4LG exhibits three types of folding, 3+3 f6LG also follows this criterion. Figure 2 presents optical and Raman images of $\theta_{medium}$ and $\theta_{large}$ f6LG together with the Raman images of the low frequency modes. The same as *r*-f4LG, in the G mode resonant region of 3+3 folded layers (*r*-f6LG), the low frequency peaks are remarkably enhanced and correlate very well with the resonant G mode as visualized by their Raman images. Therefore, the responsibility of the folding induced VHS for the enhancement of G and C peaks could be extended to the 3+3 *r*-f6LG. As predicated by the theory[25] and illustrated by the diagram (Fig. 2n), there are two shear modes in Bernal-stacked TLG, locating at relatively lower and higher frequency sides of the shear mode in BLG. The Raman spectrum (Fig. 2g) of the *r*-f6LG clearly presents the two low frequency peaks. The positions and linewdiths (Table S2) guide us to assign these two modes as the Raman-active *E''* shear mode for the lower frequency one ($C_{31}$) and the infrared (IR)/Raman-active *E'* shear mode for the higher frequency one ($C_{32}$). Though $C_{31}$ is slightly weaker than $C_{32}$, it is also significantly enhanced. This is the *first* observation of the lower frequency shear mode, which is supposed to be extremely weak and could not be observed even in bulk graphite[25]. It is noticed that the interlayer coupling strength derived from the two shear modes of 3+3 *r*-f6LG is nearly identical to the one in the 2+2 *r*-f4LG and the previously reported[25]. The peak locating at ~120 cm$^{-1}$ in the r-f6LG is attributed to the combinational mode of the higher frequency breathing mode (IR-active) and the lower frequency shear mode (Raman-active), labelled as B+$C_{31}$. Very interestingly and meaningfully, in a 2+3 folded 5-layer graphene of a medium rotational angle (11.8º), G mode and shear modes of both BLG and TLG are enhanced (Figure 2). In a 6+6 *r*-f12LG, all FIVE shear modes are clearly resolved and show perfect agreement with the theoretical predication (Figure S3 and Table S3). This immediately indicates how robust the formation of VHS in such folded graphene layers, the enhancement of shear modes and G mode, and shear modes against the folding could be.

To further demonstrate the feasibility of the enhancement of shear modes by a proper folding and exploit their more intrinsic properties, we performed polarization and *in-situ* temperature-dependent Raman spectroscopy study. Figure 3 shows the low frequency and G modes of *r*-f4LG as a function of angles between the polarization of the incident and scattering lights. The strong and sharp peak locating at ~30 cm$^{-1}$ is inert to the change of polarization configurations while the intensity of the weak peak locating at ~115 cm$^{-1}$ is maximized under the parallel polarization and minimized for the perpendicular configuration. From our discussion and the assignments above, the enhanced sharp peak in the 2+2 *r*-f4LG should be the shear mode of Bernal-stacked BLG with the symmetry of $E_g$. Thus, it is in-plane two-fold degenerated and naturally independent to our polarization configurations as G mode. For the weak peak, the assignment is the combinational modes of breathing ($A_{1g}$) and shear ($E_g$) modes in *r*-f4LG. Since the out-of-plane breathing mode ($A_{1g}$) contributes to the combinational modes, the intrinsic polarization nature of the $A_{1g}$ mode is perfectly responsible for the polar-dependence of the weak peaks locating at ~115 cm$^{-1}$, which shows zero intensity under anti-parallel and maximum for parallel polarization configuration. More discussion can be found in the supplementary information.

*In-situ* temperature-dependent Raman spectroscopy is one of the most powerful tools to probe phonons, an assembling of lattice vibration and their interaction with other particles/quasiparticles. In Figure 4, we present the evolution of the shear mode of BLG in a temperature range of 90 K to 390 K. Firstly, we compare our thermal chamber temperature readings with the sample local temperatures estimated from the intensity ratio of Stokes and anti-Stokes[32]. The fairly good agreement between each other affirms that the laser heating could be neglected (Fig. 4b). Now, we focus on the line shift as a function of temperatures. A redshift of the shear mode as the increase of temperatures is observed (Fig. 4c upper panel). In the previous studies, the similar redshift of the G mode is also reported, and the frequency of G mode at 0 K and the first-order temperature coefficient were extracted by a linear fitting[33]. Softening of phonons at a higher temperature is common for many crystals owing to the enlarged bonds length due to the thermal expansion. However, graphene is exceptional as it has a quite large negative thermal expansion, potentially leading a blueshift instead, which actually has been well probed in the previous studies[34-35]. Though SLG anchoring on a substrate might be pinned down and follow the thermal expansion of the substrate, the shearing movement of the f4LG should be much free. Thus, there must be extra contribution to the overall softening of the shear mode with the increase of temperatures. The response of phonon frequency to the temperatures is a very effect manifestation of the anharmonicity. Two effects are usually responsible to the temperature-dependent line shift: anharmonic multiple phonons coupling and crystal thermal expansion[34]. We speculate that the anharmonic multiple phonons coupling should be the main reason for the softening of shear mode phonons with the increase of temperatures. Following the previous strategy[34], we fit our experimental data by a polynomial function, which carries the total effects of lattice thermal expansion and anharmonic phonons coupling. The perfect agreement of each other confirms our speculation. The frequency of the shear mode of BLG at 0 K is extrapolated to be $\omega(0) = 32.6$ cm$^{-1}$, which is critical for many further investigations, for example probing the influence of phonon-phonon coupling on the linewidth of phonon mode as discussed below. As comparison, the linear fitting is also shown here. Apparently, the nonlinear one is much more suitable, as also employed in the previous study[35].

Now, we move to the line width. In a defect-free crystal, the intrinsic linewidth ($\gamma^{in}$) is defined by: $\gamma^{in} = \gamma^{ph-ph} + \gamma^{e-ph}$, where $\gamma^{ph-ph}$ represents the anharmonic phonon-phonon coupling and $\gamma^{e-ph}$ is from the electron-phonon interaction[36]. For $\gamma^{ph-ph}$, a possible decay channel could be one shear mode phonon splits into two acoustic phonons of the same energy and opposite momentum[25] as described by: $\gamma^{ph-ph} = \gamma^{ph-ph}(0)[1+2n(\omega_0/2)]$, where $\gamma^{ph-ph}(0)$ and $\omega_0$ is the linewidth caused by the anharmonic phonon-phonon coupling and the frequency of the shear mode at 0 K, respectively. $n(\omega_T) = 1/[\exp(\hbar\omega_T/K_BT)-1]$ is the phonon occupation number, where $\hbar\omega_T$ is the shear mode phonon energy at temperature T and $K_B$ is the Boltzmann constant[37]. Rather than fixing the phonon energy, *i.e.* 196 meV in the previous study of G mode[34], we substitute individual phonon energy of shear mode at corresponding temperature because the variation of the G phonon energy is only around 0.3% in our temperature window, whereas upto 8% is noticed for the shear mode phonon energy. EPC also contributes and even become dominant contribution to linewidth in a gapless system like graphene, graphite and metallic carbon nanotubes[13,38]. For $\gamma^{e-ph}$, it should follow: $\gamma^{e-ph} = \gamma^{e-ph}(0)[f(-\hbar\omega/2K_BT) - f(\hbar\omega/2K_BT)]$, where $\gamma^{e-ph}(0)$ is the width resulted from the EPC at 0 K and $f(x) = 1/[\exp(x)+1]$. In this work, we fitted our data by considering both the anharmonic phonon decay ($\gamma^{ph-ph}$) and the EPC ($\gamma^{e-ph}$). A fairly good agreement could be achieved (Fig. 4c). $\gamma^{ph-ph}(0)$ and $\gamma^{e-ph}(0)$ are extrapolated to be 0.02 cm-1 and 17.27 cm$^{-1}$, respectively. To further elucidate the origin of the linewidth or the phonon lifetime of the shear mode, we plot the contribution of $\gamma^{ph-ph}$ and $\gamma^{e-ph}$ in Fig. 4c. It is obvious that the $\gamma^{e-ph}$ is more dominant, especially at low temperatures. We expect a substantial EPC induced increment of linewidth of shear mode at cryogenic temperature. Such decrease of phonon lifetime could be well interpreted as, at very low temperature the occupation of conduction band near the Dirac point by the thermal excited electrons could be significantly suppressed and as a result of the creation of phonon-excited electron-hole pairs and thus their interactions (EPC) are remarkably activated, leading the broadening of Rama peaks. The large contrast of the phonon energies of the shear and G modes explains why the G mode is much broader than shear mode and a large EPC of the G mode could be preserved even at a high temperature[25].

**Discussion**

Together with previous intensive Raman scattering studies of D, G, and G' modes of graphene, our systematic studies of the low frequency interlayer shear mode, C mode of FLG complement the probing of fundamental Raman studies of carbon materials. The folding-induced VHS promotes a remarkable enhancement of the shear modes as it does on the G mode. The in-plane two-fold degenerated symmetry, the anharmonicity and EPC of the shear mode are well exploited through two-dimensional Raman mapping, polarization- and temperature-dependent Raman spectroscopy of this low frequency shear mode (~5 meV), which was far away from being accessible before. More insight understandings of mechanical and electrical properties and further developments of practical applications of FLG are expected to be achieved soon through investigations of the enhanced shear modes in the stretched, electrically or molecularly doped folded FLG and even under a magnetic field.

**Methods**

**Sample preparation**

Graphene layers were prepared by the mechanical cleavage of graphite and transferred onto a 285 nm $SiO_2$/Si substrate. During the mechanical exfoliation process, some graphene flakes flipped over and folded themselves partially and accidentally. Such interesting folded graphene layers were located under an optical microscope. The number of layers of the unfolded part was further identified by white light contrast spectra and Raman spectroscopy[39]. The folding or rotational angles were determined by reading the *R* peak position[23-24] and double checked by their geometrical morphologies visualized in their optical and Raman images[40].

**Raman spectroscopy study**

A WITec CRM200 Raman system with a low-wavenumber coupler, a 600 lines/mm grating, a piezocrystal controlled scanning stage, a ×100 objective lens of NA = 0.95 was used for the Raman images. For Raman spectra of good spectral resolution, a 2400 lines/mm grating was used. The in-situ temperature-dependent Raman measurements were conducted in a Linkam thermal stage with a ×50 objective lens of NA = 0.55. All the Raman images and spectra were recorded under an excitation laser of 532 nm ($E_{laser}$ = 2.33 eV). To avoid the laser-induced heating, laser power was kept below 0.1 mW.

**Acknowledgements**

This work is supported by the Singapore National Research Foundation under NRF RF Award No. NRF-RF2010-07 and MOE Tier 2 MOE2012-T2-2-049. C.X.C thanks the valuable discussion with Dr Jun Zhang. T.Y. thanks the enlightening discussion with Professor Castro Neto, Antonio Helio. The authors are grateful for the important help of Dr Jeil Jung on sharing the electronic band structure and density of states of 2+2 folded graphene layers.

**Author contributions**

C.X.C. and T.Y. initialled the project; conceived and designed the experiments; C.X.C. performed the experiments; C.X.C. and T.Y. analysed the data, discussed the results and co-wrote the paper.

## Additional information

Supplementary information is available in the online version of the paper. Reprints and permissions information is available online at www.nature.com/reprints. Correspondence and requests for materials should be addressed to T. Y.

## Competing financial interests

The authors declare no competing financial interests

Note: We noticed a similar work by Dr. Tan P. H. *et al.* from a conference after our submission.

**Figure captions**

**Figure 1 | Raman images and spectra of 2+2 f4LG. a**, Optical image of folded BLG. The folding types are identified and labelled by their rotational angles. Raman intensity images of **b**, G mode; **c**, G' mode; Raman image of **d**, intensity; **e**, frequency; and **f**, width of the shear mode of BLG ($C_2$). **g**, Raman spectra of low and intermediate frequency modes of BLG, $\theta_{medium}$ 2+2 *r*-f4LG and $\theta_{large}$ 2+2 f4LG. **h**, Raman spectra of low and intermediate frequency modes of 2+2 *r*-f4LG with different rotational angles as determined by the *R* peak positions and indicated. ($E_{laser}$ = 2.33 eV).

**Figure 2 | Raman images and spectra of 3+3 *r*-f6LG (left panel) and 2+3 *r*-f5LG (right panel). a,** Optical image of folded TLG. The folding types are identified and labelled by their rotational angles. Raman intensity images of **b**, G mode; **c**, G' mode; **d**, lower frequency shear mode ($C_{31}$); **e**, higher frequency shear mode ($C_{32}$), and **f**, combinational mode (B+$C_{31}$). **g**, Raman spectrum of low frequency modes of 3+3 *r*-f6LG with rotational angle of 12.2°. Raman image of **h**, intensity of G mode; **i**, intensity of G' mode; **j**, width of G' mode; **k – m**, Raman intensity images of shear modes in BLG ($C_2$) and in TLG ($C_{31}$ and $C_{32}$); **n**, schematic diagram of shear modes in BLG and TLG. **o**, Raman spectrum of low and intermediate frequency modes of 2+3 *r*-f5LG. The shear modes of both BLG and TLG are enhanced together with the G mode. ($E_{laser}$ = 2.33 eV).

**Figure 3 | Polarization-dependent Raman spectra of 2+2 *r*-f4LG.** Raman spectra of **a**, low frequency modes, and **b**, G mode when the angles between the polarization of the incident and scattering lights are tuned to be 0, 30, 60 and 90 degrees as indicated. ($E_{laser}$ = 2.33 eV).

**Figure 4 | *In-situ* temperature-dependent Raman spectra of 2+2 *r*-f4LG. a,** Temperature-dependent Raman spectra of shear mode with both Stokes and anti-Stokes lines; **b**, sample local temperatures estimated from the intensity ratio of Stokes and anti-Stokes as a function of the thermal chamber temperatures; **c**, positions (upper panel) and full width at half maximum (FWHM) (lower panel) of Stokes shear mode as a function of temperatures. The solid spheres represent the experimental data. The straight blue line (upper panel) is the linear fit and the pink curve is the polynomial fit, which accounts for both thermal expansion and anharmonic multi-phonon interaction. Obviously, the nonlinear fit is superior to the linear one. The frequency of the shear mode at 0 K can be extrapolated. For FWHM (lower panel), the green line is the fit of the data by considering both effects of the phonon-phonon (ph-ph) interaction and electron-phonon coupling (EPC). The purple and blue plots describe the contributions of the ph-ph and EPC, separately. The dominance of EPC, especially at low temperature is clear. Note: the FWHM has been corrected by subtracting the broadening of our system from the fitted values. ($E_{laser}$ = 2.33 eV).

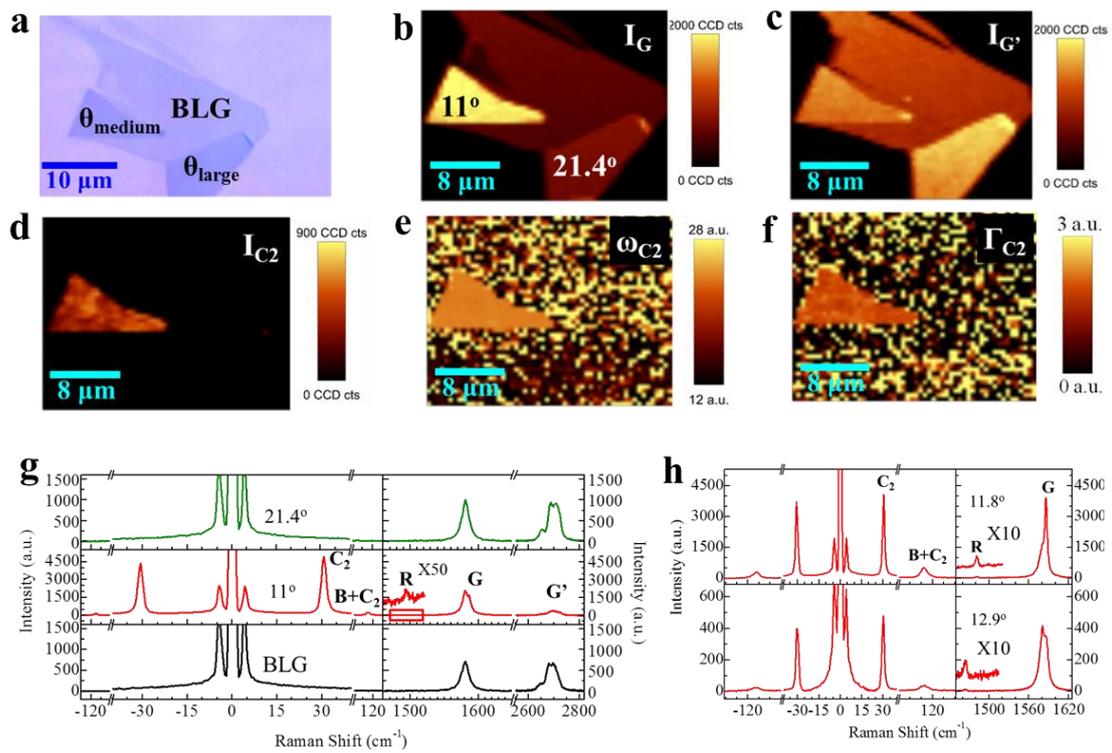

**Figure 1**

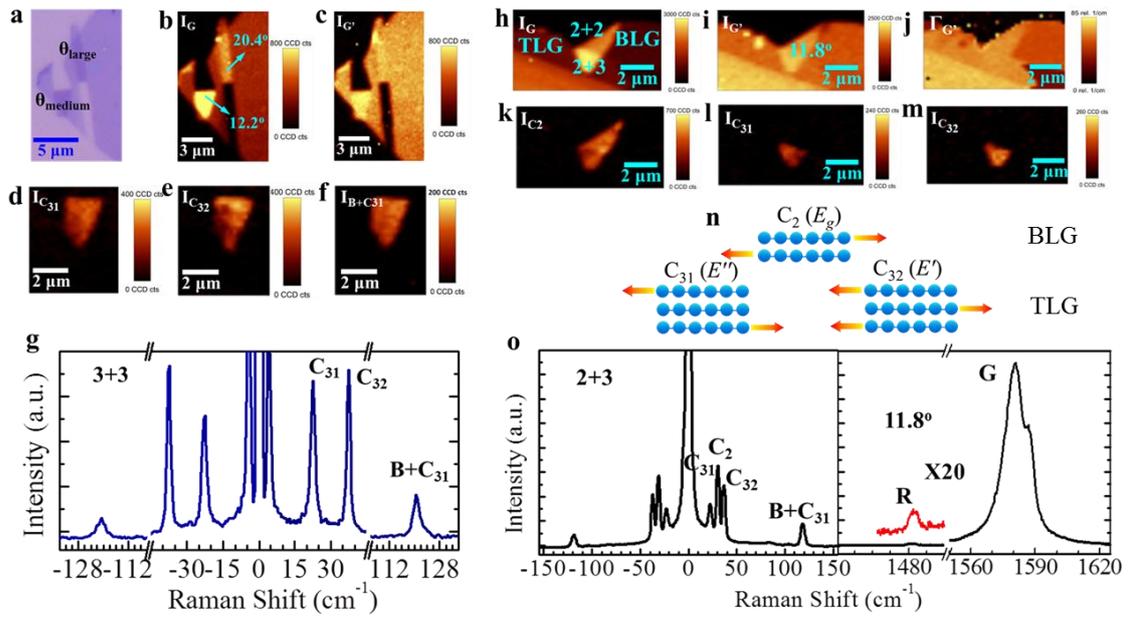

**Figure 2**

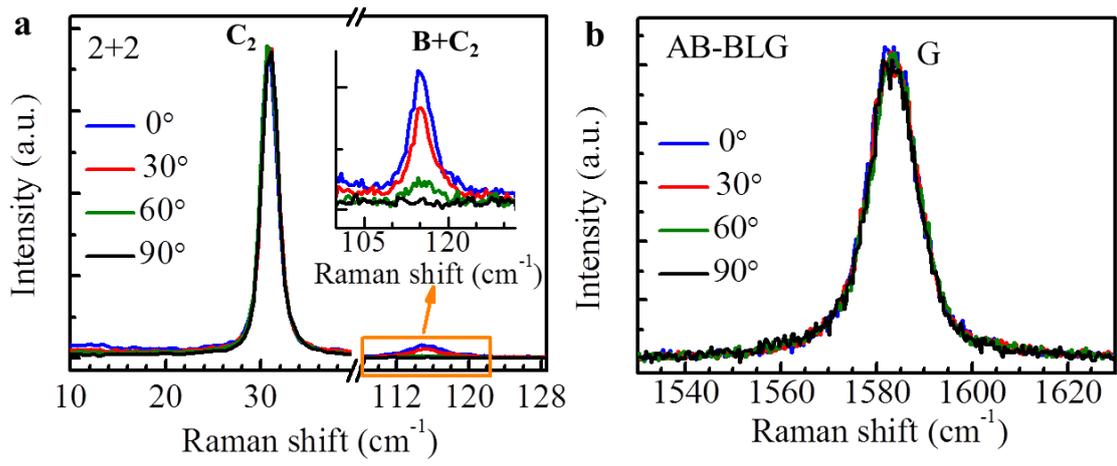

**Figure 3**

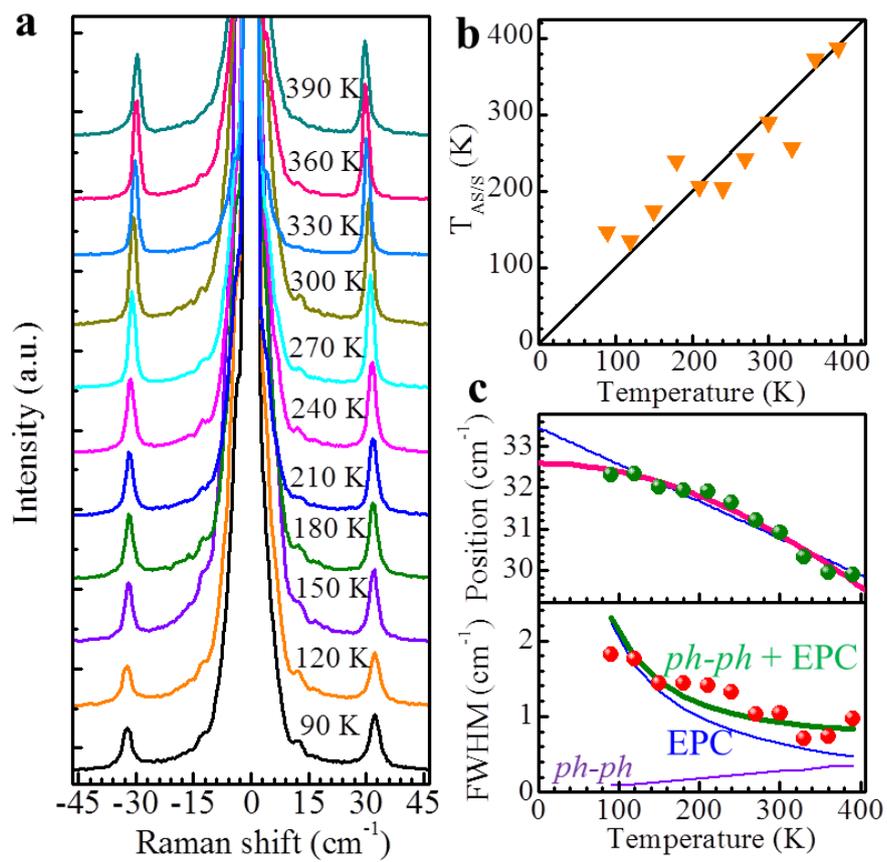

**Figure 4**

# SUPPLEMENTARY INFORMATION

## Resonant Low Frequency Interlayer Shear Modes in Folded Graphene Layers

Chunxiao Cong[1] and Ting Yu[1,2,3*]

[1]Division of Physics and Applied Physics, School of Physical and Mathematical Sciences, Nanyang Technological University, 637371, Singapore;

[2]Department of Physics, Faculty of Science, National University of Singapore, 117542, Singapore;

[3]Graphene Research Center, National University of Singapore, 117546, Singapore

*Address correspondence to yuting@ntu.edu.sg

1. **Raman spectroscopic features of low (< 130 cm$^{-1}$) frequency and intermediate (1400 – 1700 cm$^{-1}$) frequency modes in resonant folded few-layer graphene (*r*-fFLG).**

2+2 *r*-f4LG and 3+3 *r*-f6LG layers of different rotational angles are studied. From the position of *R* peaks, the rotational angles are determined[1-2]. Considering the spectral broadening of our system, ~0.9 cm$^{-1}$, we further correct the fitted line widths and list in the tables below. The interlayer coupling strength (*α*) is able to be obtained by:

$$\omega_i^2 = \frac{1}{2\pi^2 c^2}\frac{\alpha}{\mu}\left\{1-\cos\left[\frac{(i-1)\pi}{N}\right]\right\}$$, here *i* = 2, 3, …6; $\omega_i$ is the frequency of each shear mode, $\mu = 7.6 \times 10^{-27}$ kg Å$^{-2}$, is the SLG mass per unit area and $c = 3 \times 10^{10} cm s^{-1}$, is the speed of light[3].

**Supplementary Table S1 | Positions (widths) of $C_2$, $B+C_2$, and R modes in *r*-f4LG together with interlayer coupling strength (α) derived from equation above.**

| Rotational angles | $C_2$ peak Position (width) (cm$^{-1}$) | $B+C_2$ peak Position (width) (cm$^{-1}$) | R peak Position (width) (cm$^{-1}$) | interlayer coupling strength α (Nm$^{-3}$) |
|---|---|---|---|---|
| 2+2 (11°) | 30.7 (0.9) | 115.1 (3.7) | 1493.8 (4.2) | 12.7 ×10$^{18}$ |
| 2+2 (11.8°) | 30.6 (0.7) | 114.9 (3.6) | 1481.4 (4.2) | 12.6 ×10$^{18}$ |
| 2+2 (12.9°) | 30.2 (0.9) | 115.0 (4.6) | 1464.1 (4.6) | 12.3 ×10$^{18}$ |

**Supplementary Table S2 | Positions (widths) of $C_{31}$, $C_{32}$, $B+C_{31}$, and R modes in *r*-f6LG together with interlayer coupling strength (α) derived from equation above.**

| Rotational angle | C peak Position (width) (cm$^{-1}$) | $B+C_{31}$ peak Position (width) (cm$^{-1}$) | R peak Position (width) (cm$^{-1}$) | interlayer coupling strength α (Nm$^{-3}$) |
|---|---|---|---|---|
| 3+3 (12.2°) | $C_{31}$: 22.6 (1.1) | 120.6 (1.7) | 1471.8 (4.7) | 13.8 ×10$^{18}$ |
| | $C_{32}$: 37.5 (0.8) | | | 12.6 ×10$^{18}$ |

## 2. Fitting the low frequency shear mode by a single Lorentzian line shape peak.

In previous study[3], due to the quantum interference between the shear mode and a continuum of electronic transition near the K point, a Breit-Wagner-Fano (BWF) line shape is observed in the weak shear modes. In this work, the shear modes are dramatically enhanced and could be fairly well fitted by a single Lorentzian line shape peak.

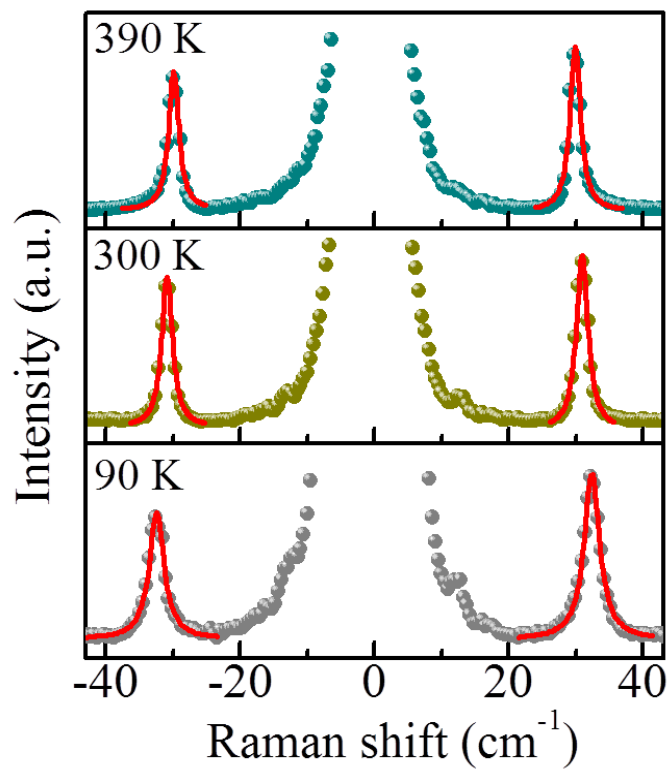

**Supplementary Figure S1 | Raman spectra of the low frequency shear mode in 2+2 *r*-f4LG obtained at 90 K, 300 K and 390 K.** Both Stokes and anti-Stokes peaks are well fitted by a single Lorentzian line shape peak.

## 3. Electronic band structure and density of states (DOS) of the 2+2 *r*-f4LG.

In our previous study[4], we grouped 2+2 folded graphene layers into three types by reading the folding or rotational angles θ, such as $θ_{small}$ (<4 degrees), $θ_{medium}$ (around 11 degrees), and $θ_{large}$ (>20 degrees), for the excitation photon energy of 2.33 eV. Here, we plot the electronic band structure and density of states (DOS) of a 2+2 f4LG with a medium rotational angle of 11.2 degree for 2.33 eV.

The electronic structure was obtained applying the methodology outlined in previous work for twisted layer systems[5-6] here for the specific system of 2+2 f4LG. The band structures are represented in the Moire Brillouin zone and are plotted along the following symmetry lines.

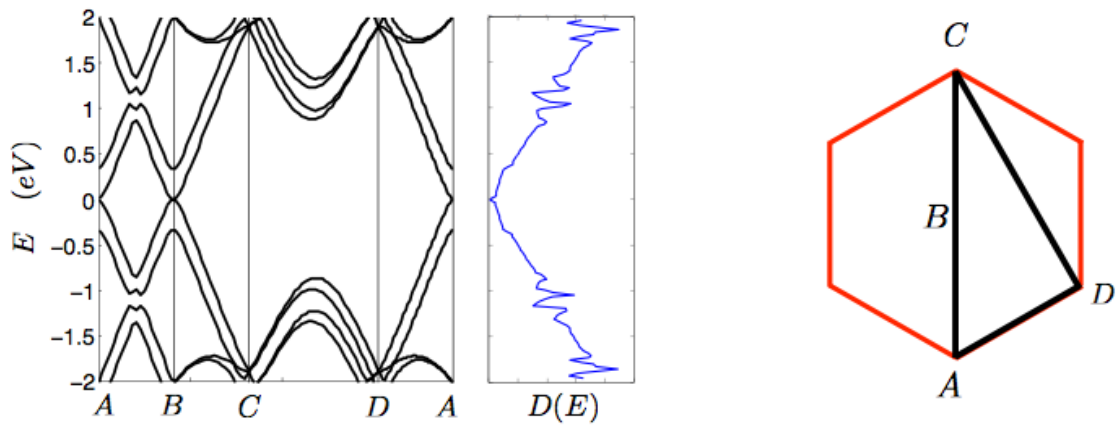

**Supplementary Figure S2 | Electronic band structure and density of states (DOS) of the 2+2 *r*-f4LG with a rotational angle of 11.2 degree.** The band structure is represented along the straight lines connecting the symmetry points of the Moire Brillouin zone represented on the right panel.

## 4. Shear modes in 6+6 *r*-f12LG.

The enhancement of shear modes of 6-layer graphene (6LG) is obtained in a 6+6 *r*-f12LG. All the shear modes are resolved and show good agreement with the theoretical prediction based on a simple yet sufficient linear-chain model as[3]:

$$\omega_i^2 = \frac{1}{2\pi^2 c^2}\frac{\alpha}{\mu}\left\{1-\cos\left[\frac{(i-1)\pi}{N}\right]\right\}$$, here $i$ = 2, 3, …6; $\omega_i$ is the frequency of each shear mode, $\alpha = 12.8\times10^{18} Nm^{-3}$ is the interlayer force constant per unit area, $\mu = 7.6\times10^{-27} kg\ Å^{-2}$ is the SLG mass per unit area and $c$ is the speed of light in $cms^{-1}$.

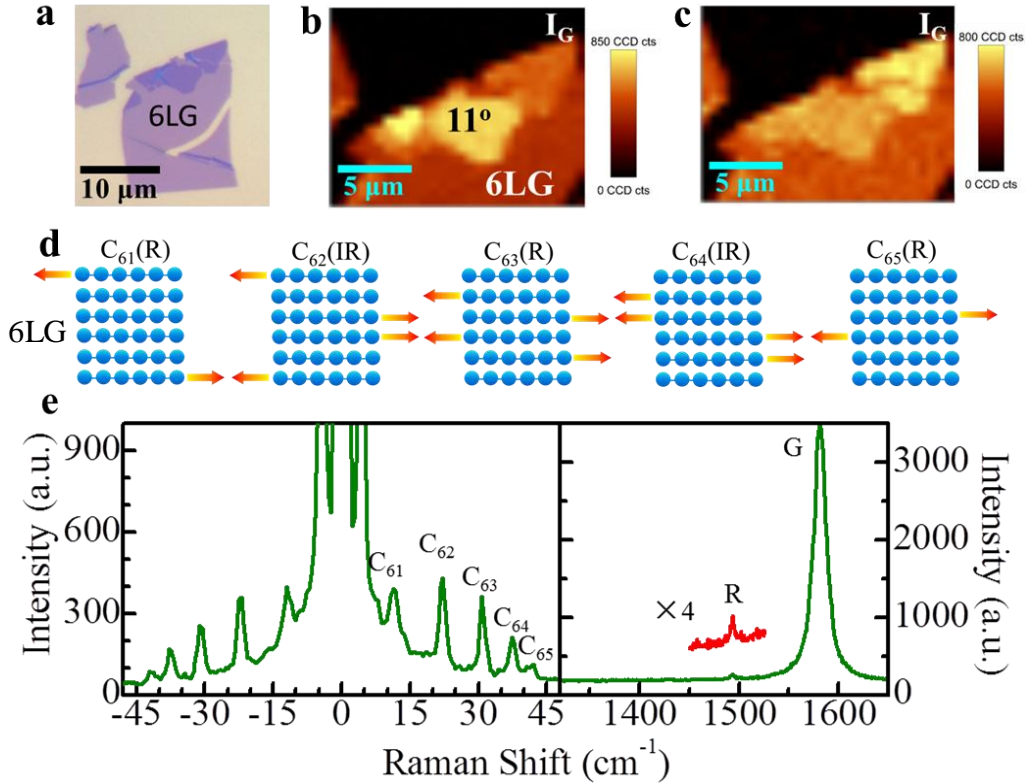

**Supplementary Figure S3 | Raman images and spectrum of 6+6 *r*-f12LG. a,** Optical image of folded 6-layer graphene; Raman intensity images of **b**, G; **c**, G' bands; **d**, Schematic diagram of shear modes of 6LG. The Raman-active (R) or IR-active (IR) are also indicated; **e**, Raman spectrum of shear, R and G modes of 6+6 *r*-f12LG. The IR-

active modes become Raman-active might be due to the folding as discussed in our previous study[4].

**Supplementary Table S3 | Positions (Stokes) and widths of $C_{61} - C_{65}$ shear modes in 6+6 *r*-f12LG together with the linear-chain mode predicted positions as a comparison.**

| Rotational angle | C peak Expt. Position (width) (cm$^{-1}$) | C peak Calculated Position (cm$^{-1}$) | R peak Position (width) (cm$^{-1}$) |
|---|---|---|---|
| 6+6 (11°) | $C_{61}$: 11.3 (2.7) | $C_{61}$: 11.3 | 1493.7 (4.7) |
|  | $C_{62}$: 22.1 (0.7) | $C_{62}$: 22.8 |  |
|  | $C_{63}$: 30.9 (0.7) | $C_{63}$: 30.8 |  |
|  | $C_{64}$: 37.4 (1.0) | $C_{64}$: 37.7 |  |
|  | $C_{65}$: 41.6 (1.3) | $C_{65}$: 42.1 |  |

# 5. Determination of the polarization dependence of shear modes and their combinational modes in *r*-f4LG.

Raman mode intensity is proportional to $|e_i \cdot \mathfrak{R} \cdot e_s|^2$ where $e_i$ and $e_s$ are the unit vectors describing the polarizations of the incident and scattered lights, and $\mathfrak{R}$ is Raman tensor. In this work, the polarization of the incident light is fixed along the horizontal ($e_i = \begin{bmatrix} 1 \\ 0 \\ 0 \end{bmatrix}$) while the polarization of the scattered light is tuned with an angle ($\varphi$) to the horizontal by a polarizer ($e_s = \begin{bmatrix} \cos\varphi \\ \sin\varphi \\ 0 \end{bmatrix}$). The Raman tensors for the shear and the breathing modes of BLG are:

shear mode $E_g = \begin{bmatrix} c & 0 & 0 \\ 0 & -c & 0 \\ 0 & 0 & 0 \end{bmatrix}, \begin{bmatrix} 0 & d & 0 \\ d & 0 & 0 \\ 0 & 0 & 0 \end{bmatrix}$ ($c = d$) and breathing mode $A_{1g} = \begin{bmatrix} a & 0 & 0 \\ 0 & a & 0 \\ 0 & 0 & b \end{bmatrix}$.

Thus, the intensity of the shear mode is: $I_C \propto c^2$ and the intensity of the breathing mode is: $I_B \propto a^2 \cos^2 \varphi$.

# 6. Estimation of sample local temperatures from the intensity ratio of Stokes and anti-Stokes peaks of shear mode.

The sample local temperatures are estimated by reading the intensity ratio of Stokes ($I_S$) and anti-Stokes ($I_{AS}$) lines as[7]:

$$\frac{I_{AS}}{I_S} = \left(\frac{\omega_L + \omega_C}{\omega_L - \omega_C}\right)^4 \exp\left(\frac{\hbar \omega_C}{K_B T}\right)$$

where $\omega_L$ represents the laser frequency, $\omega_C$ represents the shear mode (C peak) frequency, $K_B$ is the Boltzmann constant and $T$ is the temperature.